\documentclass[seceq,epsf]{ptptex}

\usepackage{graphicx}
\usepackage{wrapft}

\newcommand{\SUZAKU}{{\it Suzaku}}

\newcommand{\EINSTEIN}{{\it Einstein}}

\newcommand{\XMM}{{\it XMM-Newton}}

\newcommand{\CHANDRA}{{\it Chandra}}

\newcommand{\UNITLUMI}{{\rm ergs~s$^{-1}$}}

\newcommand{\UNITNH}{{\rm cm$^{-2}$}}

\newcommand{\NH}{{\it N$_{\rm H}$}}

\newcommand{\LX}{{\it L$_{\rm X}$}}

\newcommand{\KT}{{\it kT}}

\newcommand{\etacar}{$\eta$~Car}

\title{
Diffuse X-ray Emission from the Carina Nebula Observed with Suzaku%
}

\author{
Kenji \textsc{Hamaguchi}$^{1,2}$, the Suzaku $\eta$ Carinae team and the Carinae D-1 team%
}

\inst{
$^{1}$CRESST and X-ray Astrophysics Laboratory NASA/GSFC,
Greenbelt, MD 20771\\
and\\
$^{2}$Universities Space Research Association, 
10211 Wincopin Circle, Suite 500, Columbia, MD 21044
}

\abst{
A number of giant HII regions are associated with soft diffuse X-ray emission. Among these, the Carina nebula possesses the brightest soft diffuse emission. The required plasma temperature and thermal energy can be produced by collisions or termination of fast winds from main-sequence or embedded young O stars, but the extended emission is often observed from regions apart from massive stellar clusters. The origin of the X-ray emission is unknown.

The XIS CCD camera onboard Suzaku has the best spectral resolution for extended soft sources so far, and is therefore capable of measuring key emission lines in the soft band. Suzaku observed the core and the eastern side of the Carina nebula (Car-D1) in 2005 Aug and 2006 June, respectively. Spectra of the south part of the core and Car-D1 similarly showed strong L-shell lines of iron ions and K-shell lines of silicon ions, while in the north of the core these lines were much weaker. Fitting the spectra with an absorbed thin-thermal plasma model showed \KT$\sim$0.2, 0.6~keV and \NH$\sim$1$-$2$\times$10$^{21}$~\UNITNH\ with a factor of 2-3 abundance variation in oxygen, magnesium, silicon and iron. The plasma might originate from an old supernova, or a super shell of multiple supernovae.
}

\begin{document}

\maketitle

\section{Extended X-ray Emission from the Star Forming Region}
Soft X-ray emission nebulae with \KT $\sim$0.1--0.8~keV, log \LX $\sim$33-35~\UNITLUMI, and size of $\sim$1--10$^{3}$~pc accompany a number of giant HII regions
(see Table~4 of Ref. \citen{Townsley2003}.
{\it Chandra} observations of extended emission in a few star forming clusters indicate that the emission may arise from the fast O star stellar winds thermalized either 
by wind-wind collisions or by a termination shock.
However, the emission is often found outside of the massive stellar clusters, so that another origin, such as an otherwise unrecognized supernova remnant, cannot be ruled out.

In principle, the origin of the diffuse emission can be determined by measuring its composition.
For example, the plasma should be overabundant in nitrogen and neon if it originates from
winds from nitrogen-rich Wolf-Rayet stars (WN),
while it would be overabundant in oxygen if it arises from a Type~II SNR.
The temperature of the plasma, typically a few million degrees, makes soft X-ray band studies
highly desirable, because of the presence in this band of strong lines from these elements, plus
carbon, silicon and iron.

The Carina Nebula, which contains several evolved and main-sequence massive stars
such as \etacar, WR~25 and massive stellar clusters such as Trumpler~14 (Tr 14),
emits soft diffuse X-rays 10--100 times stronger
than any other Galactic giant HII region 
(\LX\  $\sim$10$^{35}$~\UNITLUMI) \cite{Seward1982}.
The high surface brightness made possible the discovery of the diffuse emission by
the \EINSTEIN\  Observatory in the late 1970's.
The \EINSTEIN\  observations revealed that
the diffuse emission tends to be associated with optically bright regions containing massive stars.
Recent \CHANDRA\  observations provided a point source free measurement of the diffuse flux  
\cite{Evans2003}, and suggested the
presence of a north-south Fe and Ne abundance gradient \cite{Townsley2006}.

The X-ray CCD cameras (XISs: X-ray Imaging Spectrometer) onboard the \SUZAKU\  observatory
have the best spectral resolution for extended soft X-ray emission and thus they provide good
diagnostics of emission lines especially below $\sim$1~keV.

\section{{\it Suzaku} and {\it XMM-Newton} Observations of the Carina Nebula}

\begin{figure}[t]
\centerline {
\includegraphics[width=13.5cm]{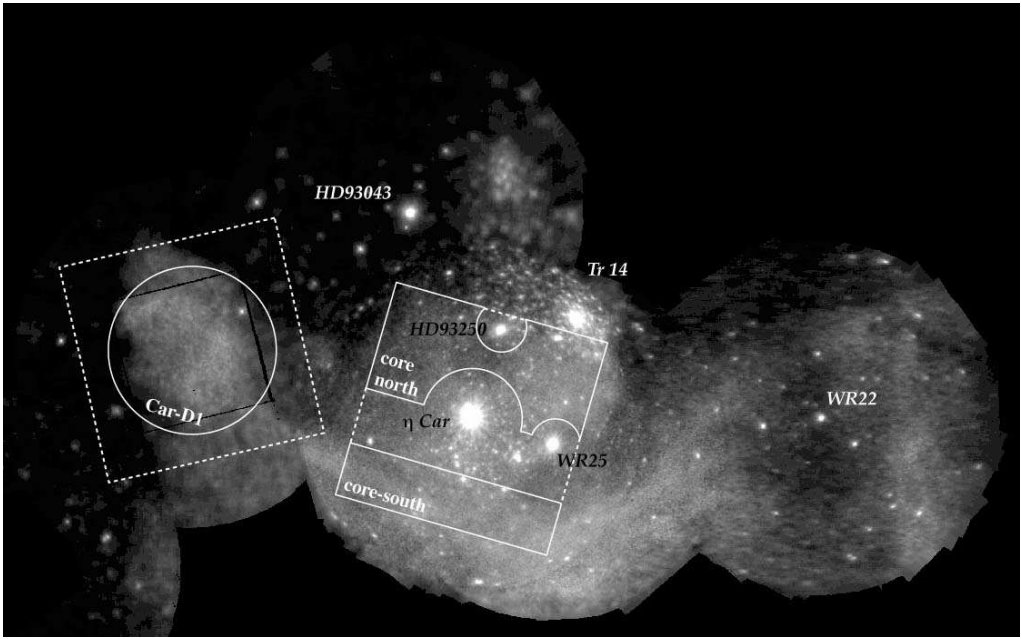}
}
\caption{Mosaic image ($\sim$90$'\times$60$'$) of the Carina nebula between 0.4$-$7~keV created 
from 32 \XMM\ observations.
The image is created with the ESAS package,
divided by the exposure map and smoothed with the adaptive smoothing technique.
The dotted lines show the XIS FOVs of the \SUZAKU\ observations of \etacar\ ({\it right})
and the Car-D1 field ({\it left}).
The solid lines show source extraction regions for the spectral analysis.}
\label{fig:carmosaicimg}
\end{figure}

Figure~\ref{fig:carmosaicimg} shows a mosaic image of the Carina nebula between 
0.4$-$7~keV created from 32 \XMM\ observations.
The image depicts several bright X-ray point sources:
\etacar\  (an LBV), WR25, WR22 (Wolf-Rayet stars), HD~93250, HD~93043 (O3 stars),
and Tr~14, Tr~16 (massive stellar clusters).
The image also clearly shows apparently extended emission toward the east-west direction.
In a color image (e.g. Figure~1 of Ref. \citen{Hamaguchi2007a}, 
\XMM\ Image Gallery\footnote{http://xmm.esac.esa.int/external/xmm\_science/gallery/public})
the emission is softer between Tr~14, WR~25 and \etacar.

We analyzed the \SUZAKU\ data of the core and the eastern side (named Car-D1)
of the Carina nebula taken on 2005 Aug. 29 and 2006 June 5.
The XIS FOVs of these observations are shown in Figure~\ref{fig:carmosaicimg} with dotted lines.
To investigate the color variation in detail,
we divided the core region into two and thus extracted three spectra from two \SUZAKU\ observations
(core-north, core-south and Car-D1).
The background was reproduced with the night earth data.
The spectra showed strong emission between 0.3 and 2~keV, which is probably dominated by
soft diffuse emission associated with the Carina nebula, while
the spectra above 2 keV may be explained with
CXB, Galactic Ridge X-ray Emission, X-ray point sources resolved with \CHANDRA\ 
and unresolved pre-main-sequence stars.

\begin{figure}[t]
\centerline {
\includegraphics[width=6.5cm]{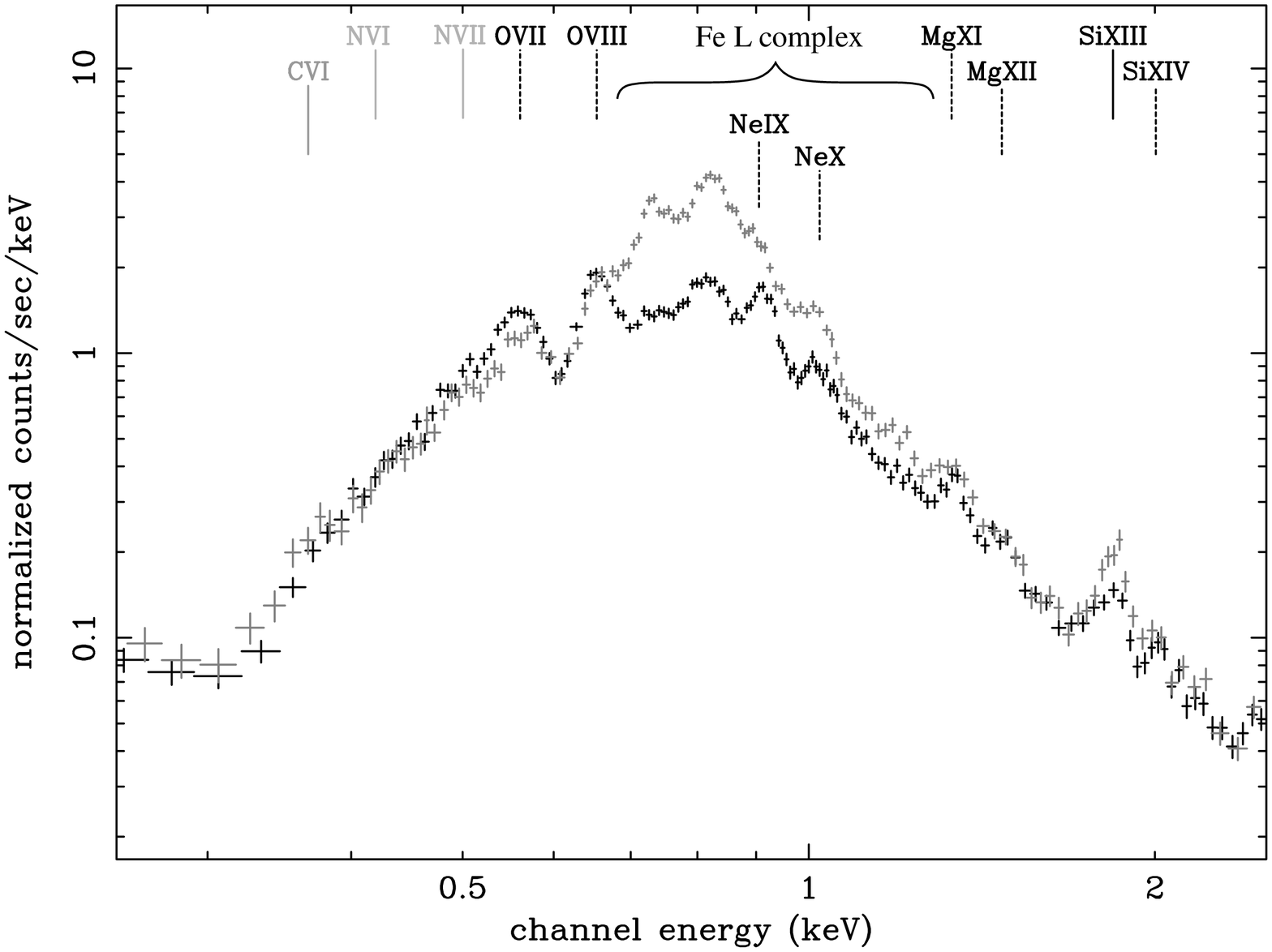}
\includegraphics[width=6.5cm]{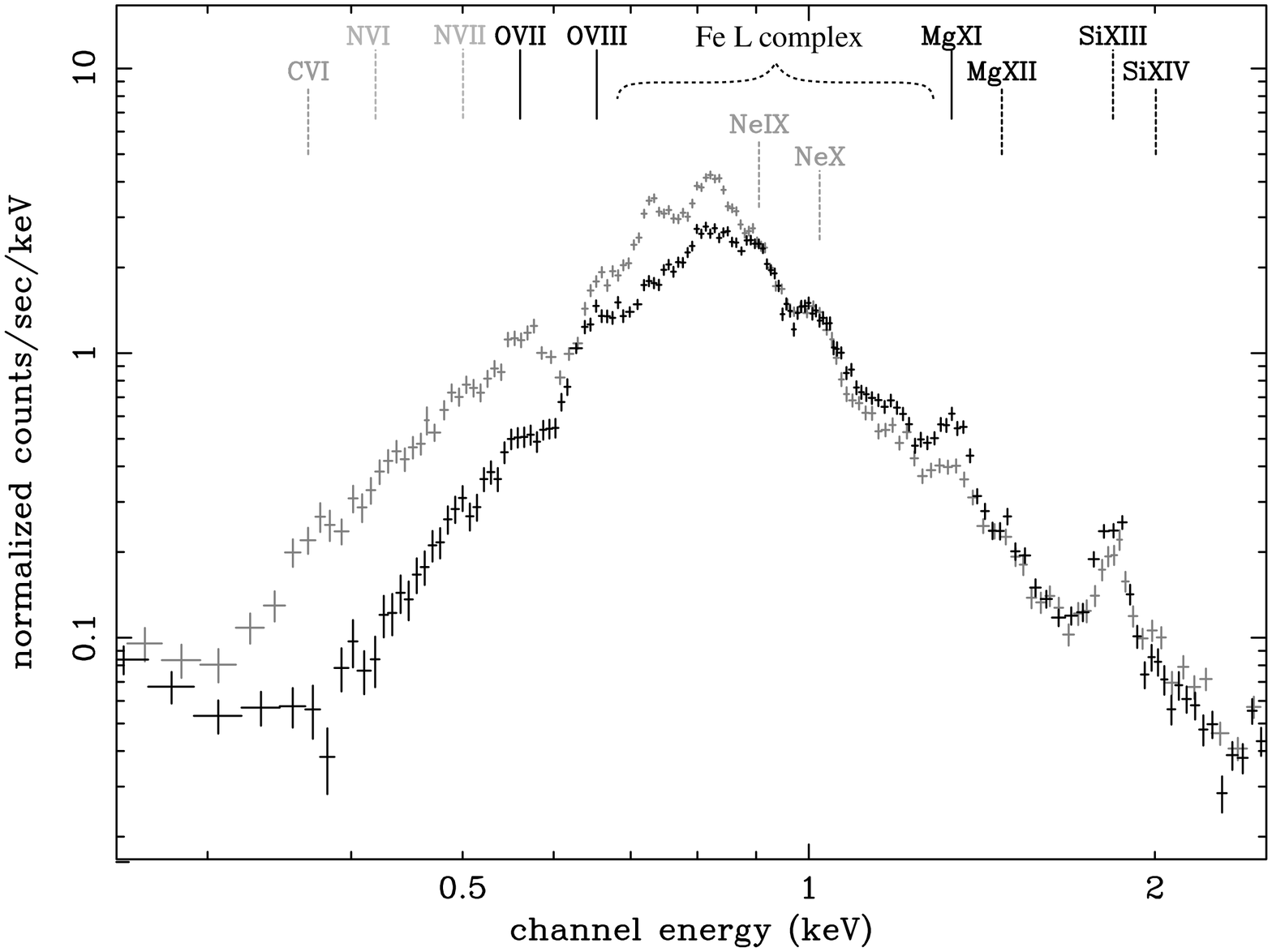}
}
\caption{
Comparison of the XIS1 spectra between the fields -- {\it left}: the core-north region (black) and
the core-south region (grey), {\it right}: the Car-D1 field (black) and
the core-south region (grey).
The above labels demonstrate energies of emission lines detected (black) 
or concerned (grey) with this result.
Emission lines with the solid lines showed variation in their line intensity.
Low count rates of the Car-D1 spectrum below 1~keV is caused by degradation
of soft response by progressive contamination on the XIS.
}
\label{fig:carspec}
\end{figure}

Figure~\ref{fig:carspec} shows an overlay of the BI spectra between 0.3--2~keV.
The left panel compares spectra of the core-north region with the core-south region.
A strong difference is seen between 0.7~keV and 1.2~keV, which
apparently is the source of the two colors of diffuse emission.
The band in which the difference is found is dominated by emission lines from the iron L-shell complex.
Additionally, the core-south spectrum shows a stronger Si line.
The Car-D1 spectrum shows similar intensity in the Si and Fe lines to the core-south spectrum 
(right panel of Figure~\ref{fig:carspec})
while it shows relatively strong magnesium and oxygen lines.
All these spectra look similar except for these emission lines.
This suggests that the differences represent an
elemental abundance variation, and not a temperature difference.

This is supported by spectral fits of the individual spectra.
All three spectra between 0.3$-$2~keV were reproduced by an absorbed
2T thin-thermal plasma models although the best-fit models are not formally acceptable.
The plasma temperatures of all three regions are $\sim$0.2 and $\sim$0.6~keV,
and their column densities are $\sim$3$\times$10$^{21}$~\UNITNH,
which is consistent with extinction toward the Carina nebula \cite{Leutenegger2003}.
The abundances of some elements show a factor of 2$-$4 variations:
the core-north region has a factor of 2 lower silicon abundance and a factor of 4 lower
iron abundance than the core-south region,
while the Car-D1 region has a factor of 2 higher oxygen and magnesium abundances.
On the other hand,
spectral fits of the core region with higher sensitivity around 0.5~keV 
gave small upper-limits ($\lesssim$0.02~solar) of the nitrogen abundance.

\section{Origin of the Diffuse Plasma}

The N/O abundance ratio inferred from the spectral fits is $\lesssim$0.4,
over 20 times less than around \etacar.
The abundance distribution is totally contrary to that expected from stellar winds 
from evolved massive stars,
unless the winds somehow heat the interstellar matter without enriching it, thus leaving the 
X-ray plasma with abundances typical of interstellar matter.
At the same time, the
X-ray luminosity of the Carina Nebula is about two orders of magnitude higher than that
of other Galactic star forming regions, but the number of early O stars is only an order of magnitude higher
(see Table~4 in Ref. \citen{Townsley2003}.
These results suggest an additional energy source is needed to power the X-ray emission in the Carina Nebula.

An obvious possibility is one or more core-collapse supernovae (i.e. Type Ib,c or II), 
mentioned as a possibility by Ref. \citen{Townsley2003}.
The regions vary strongly in oxygen, magnesium, silicon, and iron abundances.
These elements are products of core-collapse supernovae, and
young SNRs such as Cas A and Vela show strong abundance variation from location 
to location.
The total energy content in the hot gas of $\sim$2$\times$10$^{50}$ ergs is a modest fraction of the $\sim$10$^{51}$ ergs of kinetic energy produced by a canonical supernova,
while assuming an iron abundance of 0.30~solar,
the total iron mass in the diffuse gas requires at least 3-5 supernovae.

\section*{Acknowledgements}
K.\, H. is financially supported by a US \CHANDRA\  grant No. GO3-4008A
and US \SUZAKU\ grant.

\end{document}